\documentclass[doublecol]{epl2} 
\usepackage{graphicx}
\usepackage{amsfonts}
\usepackage{amssymb}
\usepackage{bm}
\usepackage{color}
\usepackage{multirow}
\usepackage{times}

\title{     Compass-Heisenberg model on the square lattice \\ 
            --- spin order and elementary excitations}
\shorttitle{Compass-Heisenberg model 
            --- spin order and elementary excitations}

\author {     Fabien Trousselet\inst{1} \and Andrzej M. Ole\'{s}\inst{1,2} 
         \and Peter Horsch\inst{1} }
\shortauthor{F. Trousselet, A. M. Ole\'{s} and P. Horsch}

\institute{
  \inst{1} Max-Planck-Institut f\"ur Festk\"orperforschung,
           Heisenbergstrasse 1, D-70569 Stuttgart, Germany \\
  \inst{2} Marian Smoluchowski Institute of Physics, Jagellonian
           University, Reymonta 4, PL-30059 Krak\'ow, Poland
}


\abstract{
We explore the physics of the anisotropic compass model under the
influence of perturbing Heisenberg interactions and present 
the phase diagram with multiple quantum phase transitions. 
The macroscopic ground state degeneracy of the compass model
is lifted in the thermodynamic 
limit already by infinitesimal Heisenberg coupling, which selects
different ground states with $\mathbb{Z}_2$ symmetry depending on the sign
and size of the coupling constants --- then low energy excitations 
are spin waves, while the compass states reflecting columnar order 
are separated from them by a macroscopic gap. Nevertheless, nanoscale 
structures relevant for quantum computation purposes may be tuned 
such that the compass states are the lowest energy excitations,
thereby avoiding decoherence, if a size criterion derived by us is 
fulfilled.\\
{\it Published in: EPL {\bf 91}, 40005 (2010).}
}

\begin{document}

\maketitle

Strongly correlated electrons in transition metal oxides lead to 
rich quantum physics controlled by spin and orbital superexchange 
interactions that are complex and often intrinsically frustrated due 
to competing interactions \cite{Kug82,Fei97}. 
A high frustratedness is realized in the orbital compass model 
\cite{Kho03,Nus04,Mis04,Dou05,NuFra}, resulting in a large degeneracy 
of ground states. In contrast to SU(2) Heisenberg 
interactions which are not frustrated and isotropic in 
spin space, compass interactions are locally Ising-like but
the spin component involved in the interactions
depends on the bond
orientation. Recent interest in the compass model was
triggered by the observations that it has an interdisciplinary
character and plays an important role in a variety of contexts. 
It could: ($i$) serve as an effective model for protected qubits 
realized by Josephson arrays \cite{Dou05}, or 
($ii$) describe polar molecules in optical lattices and systems 
of trapped ions \cite{Mil07}. 
First experimental successes in constructing
special networks of Josephson junctions guided by the
compass model have already been reported \cite{Gla09}.

Materials with large spin-orbit coupling may give rise to compass
spin interactions for some lattice structures \cite{Jac09},
leading either to the compass or the Kitaev honeycomb model 
\cite{Kit06}. Numerical studies \cite{Dor05,Wen08} and the
mean-field approach \cite{Che07} suggest that when anisotropic
interactions are varied through the isotropic point of the
two-dimensional (2D) compass model, a quantum phase transition
(QPT) between two different types of directional order occurs.
Recently the existence of this transition, similar to the one
found in the exact solution of the one-dimensional (1D) compass 
model \cite{Brz07}, was confirmed using projected entangled-pair 
state algorithm \cite{Oru09}. This implies that the symmetry is
spontaneously broken at the compass point, and the spin orientation 
follows one of two equivalent interactions, as concluded recently 
within the multiscale entanglement renormalization ansatz (MERA)
\cite{Cin10}.

\begin{figure}[t!]
\begin{center}
\includegraphics[width=8.4cm]{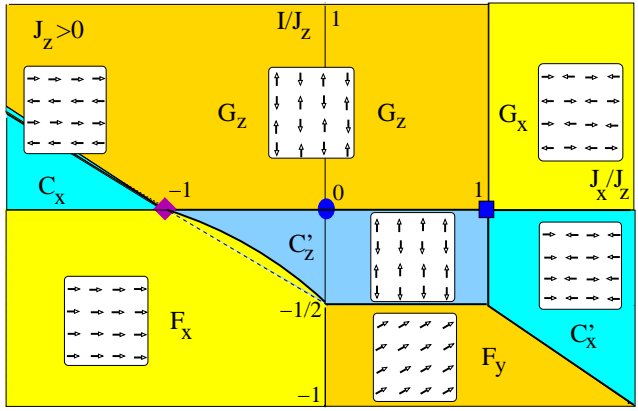}
\end{center}
\caption{Phase diagram of the CH model Eq. (1) 
with AF compass coupling $J_z>1$.
Square and diamond mark the isotropic compass
points $J_x=\pm J_z$, where in each case four ordered phases meet.
For a phase $\Phi_\alpha=G_z, C'_z, F_x, \cdots$, $\Phi$ denotes
the spin order depicted in the corresponding inset (i.e., $F$ for
a FM phase, $G$ or $C$ for AF phases) and $\alpha=x,y,z$ the spin
orientation. The QPT between $F_x$ and $C'_z$ phases (solid line) is 
affected by quantum corrections to the classical transition (dashed 
line), see text.} \label{fig:phd}
\end{figure}

In this Letter we 
introduce a generalized 2D Compass-Heisenberg (CH) model and
investigate to what extent the degenerate ground
states of the quantum compass model are robust with respect to 
perturbing interactions which may occur due to an imperfect design of 
the system. General perturbations could therefore prohibit the 
conservation of non-local quantities characteristic of the compass 
model \cite{Dor05,Dou05,Brz10}. We assume that such 
interactions are of Heisenberg type, as suggested by possible solid 
state applications \cite{Jac09,Kha01}. We have found that the compass 
ground state is fragile and an infinitesimal Heisenberg 
coupling is sufficient to lift its semi-macroscopic (exponential in 
linear size) degeneracy and to generate magnetic long range order,
either ferromagnetic (FM) or antiferromagnetic (AF) one,
which privileges a pair of columnar compass states as the ground state, 
while the other compass states 
survive as finite energy excitations. Indeed, the phase diagram
of the CH model, shown in Fig.~\ref{fig:phd}, 
appears to be very rich and exhibits different QPTs between various 
phases of $\mathbb{Z}_2$ symmetry \cite{Wen} triggered via softening 
either of spin waves or of a semi-macroscopic number of quantum states.

We consider a CH model of spins $1/2$ on the square lattice 
\begin{eqnarray}
{\cal H}&=& \sum_{i,j} \left( J_x\sigma^x_{i,j} \sigma^x_{i,j+1}
        + J_z\sigma^z_{i,j} \sigma^z_{i+1,j}\right) \nonumber \\
 &+& I\sum_{i,j}\big( \vec{\sigma}_{i,j}\cdot\vec{\sigma}_{i,j+1}
             +\vec{\sigma}_{i,j}\cdot\vec{\sigma}_{i+1,j}\big)\,,
\label{ham}
\end{eqnarray}
with $\{\sigma^\alpha_{i,j}\}$ being Pauli matrices,
$\vec{\sigma}_{i,j}=\{\sigma^x_{i,j},\sigma^y_{i,j},\sigma^z_{i,j}\}$, 
and two types of nearest neighbour interactions along bonds in the 2D 
lattice: 
($i$) the frustrated compass interactions 
$\{J_x,J_z\}$ which couple $\{\sigma^\alpha_{i,j}\}$ components along 
$\alpha$-oriented bonds (the two axes are labelled $\alpha=x,z$), and 
($ii$) Heisenberg interactions of amplitude $I$. 
We will use hereafter $\phi$ and $J_c$ to define interaction 
parameters such that 
\begin{equation}
J_z\equiv J_c\cos \phi\,, \hskip 1cm
J_x\equiv J_c\sin \phi\,,
\end{equation}
with $J_c\equiv 1$ serving as a unit.
Numerical results are obtained with Lanczos exact diagonalizations for 
periodic 2D rectangular clusters with even number of sites of up to $N=36$
($N=L_x \times L_z$ with $L_x$ and $L_z$ longitudinal dimensions, 
except for $N=18$ and $N=32$ clusters which are rotated, see Ref. 
\cite{Dag94}). Lattice symmetries (translations and $C_{2v}$ point 
group) and the $\sigma^z_{\bf{r}} \rightarrow -\sigma^z_{\bf r}$ symmetry 
(for any ${\bf r}=(i,j)$) are used to reduce the size of the Hilbert space; 
note that in contrast to the Heisenberg model $\sigma^z_{\rm
  tot}=\sum_i\sigma^z_{\bf r}$ is not conserved, but only the parity
$P=(-1)^{\sigma^z_{\rm tot}/2}$.

\begin{figure}[t!]
\begin{center}
\includegraphics[width=8.6cm]{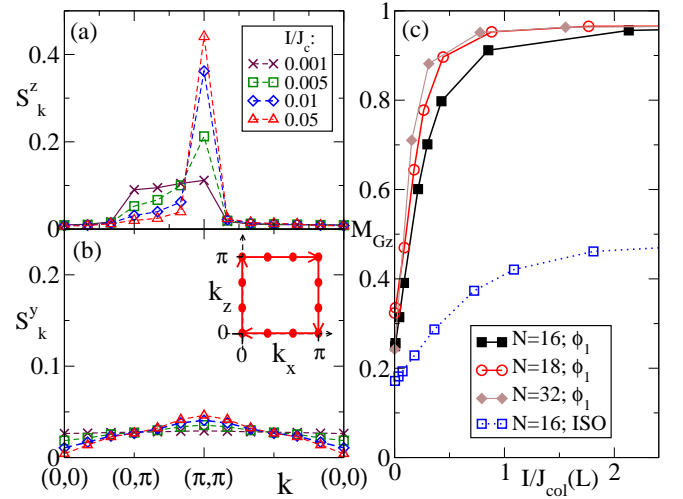}
\end{center}
\caption{Spin structure factors (a)
$S^z_{\bf{k}}$, and (b) $S^y_{\bf{k}}$, for the $N=36$-site
periodic cluster ($L_x=L_z=6$) with isotropic AF compass
couplings ($J_x=J_z$) for selected 
${\bf k}=(k_x,k_z)$, see inset in (b), and different $I/J_c$. 
(c): Order parameter $M_{Gz}$ for
cluster sizes $N=16, 18, 32$ for increasing $I/J_{\rm 
col}(L)$ (\ref{Jcol}). Solid (dotted) lines for an anisotropy
parameter $\phi_1 = \tan^{-1}(J_x/J_z)=3\pi/20$ and the isotropic 
compass model (ISO) $\phi=\pi/4$.} \label{syzq}
\end{figure}

We first examine the effect of Heisenberg interactions $I> 0$ favouring 
$G_z$-type (N\'eel) AF order  (Fig. \ref{fig:phd}). For AF anisotropic 
compass couplings $J_z>|J_x|$ spins orient along
the $z$ axis and form AF columnar states that are subsequently linked
by positive $I$ into the $G_z$-AF structure. 
The spin structure factor 
\begin{equation}
S^z_{\bf k}\equiv\frac{1}{N}\sum_{\bf r}
e^{i\bf{k}\cdot\bf{r}}\langle\sigma_{\bf 0}^z\sigma_{\bf r}^z\rangle
\end{equation}
(normalized so that $S^z_{\bf{k}}=\delta_{\bf{k},\bf{Q}}$ with
$\bf{Q}=(\pi,\pi)$ for a N\'eel state)
probes the onset of 2D spin order expected for increasing $I/J_c$, 
see Fig. \ref{syzq}(a). The compass regime with columns uncorrelated
between one another, indicated by $S^z_{\bf{k}}$ being maximal at
$k_z=\pi$ and independent of $k_x$ \cite{noteI}, 
changes surprisingly fast into the $G_z$-AF order --- the expected 
peak at $\bf{k}=\bf{Q}$ grows rapidly with increasing $I/J_c$ and is 
already of the order of unity for $I/J_c\simeq 0.01$ ! In contrast, the
peaks at $\bf{k}=\bf{Q}$ in spin structure factors $S^y_{\bf{k}}$ 
[see Fig. \ref{syzq}(b)] and $S^x_{\bf{k}}$ (not shown) grow slowly
with $I/J_c$ and are much smaller than
$S^z_{\bf Q}$ in the range of $0<I/J_c < 1$, which reflects 
the spin anisotropy and $\mathbb{Z}_2$ symmetry of the ordered $G_z$ 
phase; eventually this anisotropy weakens for increasing $I/J_c$.

The onset of the $G_z$-AF phase for $I \ll J_c$ can be understood 
using perturbation theory for small both $J_x/J_z$ and $I/J_z$. 
The unperturbed Hamiltonian 
(1D $J_z$ couplings) selects a manifold of $2^{L_x}$ column-ordered 
ground states, each column $j=1,\cdots,L_x$ possessing a $S=1/2$ 
degree of freedom is represented here by $\tau_j^z= \pm 1$, 
depending on the orientation of a reference spin in the column.

In L'th order two neighbouring columns are flipped via the horizontal 
compass couplings 
\begin{equation}
\label{Hcol}
H_{\rm col}=J_{\rm col}(L)\sum_j \tau_j^x \tau_{j+1}^x\,,
\end{equation}
with a coupling constant
\begin{equation}
\label{Jcol} J_{\rm col}(L)=8 J_z c_L (J_x/8J_z)^L\,,
\end{equation}
which gets exponentially small with increasing column length
$L\equiv L_z$ for finite systems considered here. Here $c_L$ is a 
constant increasing with $L$ ($c_L=20$, 252, 3432 for $L=4$, 6, 8 
\cite{Dor05}). $J_{\rm col}(L)$ vanishes in the thermodynamic limit 
for $J_x/J_z\leq 1$, explaining the (at least $2^{L_x}$-fold) 
ground state degeneracy of the compass model. In contrast,
Heisenberg interactions couple the $\{\sigma^z_{i,j},\sigma^z_{i,j+1}\}$ 
components in neighbouring columns, and this results in a first order
perturbative coupling $L_zI\tau_j^z\tau_{j+1}^z$ which favours 
an AF arrangement between the columns $j$ and $(j+1)$. 
This second term obviously dominates the first one ($J_{\rm col}$) 
in the thermodynamic limit as soon as $I\ne 0$, which thus ensures 
the onset of the $G_z$-AF long range order. 
This is indeed confirmed by the evolution of the order parameter
$M_{Gz}\equiv S^z_{(\pi,\pi)}$ as a function of $I/J_{\rm col}$ shown
in Fig. \ref{syzq}(c) --- while $J_{\rm col}$ varies over several
orders of magnitude between clusters considered, the
values of $M_{Gz}$ are almost identical for a given ratio $J_x/J_z$.

The isotropic ($J_x=J_z$) case is specific, as $S^z_{\bf Q}$ saturates 
then close to $\frac12$ (rather than close to 1) for 
$J_{\rm col}\ll I< J_c$, and so does $S^x_{\bf Q}$. 
In this case the compass interactions favour a manifold
of $(2^{L_x}+2^{L_z})$-degenerate ground states (in the thermodynamic 
limit), among which row-ordered states with spins along $x$ (along with
column-ordered states considered above);
the perturbing Heisenberg interactions select in this manifold the
four N\'eel states with ordered spin components either $\sigma^z_{\bf r}$
($G_z$ phase favoured for $J_z\ge J_x$) or $\sigma^x_{\bf r}$ ($G_x$
phase for $J_x\ge J_z)$. This symmetry breaking was recently studied 
in the compass model (at $I=0$) using the MERA \cite{Cin10}.

The perturbative treatment allows us to identify various ordered 
phases which develop from the $I=0$ line in the phase diagram of 
Fig. \ref{fig:phd}. For instance, the $C'_z$ phase 
(with order parameter $M_C\equiv S^z_{(0,\pi)}$) is found when the
dominant compass coupling $J_z>0$ favours AF-ordered columns while
a small $I<0$ couples neighbouring columns ferromagnetically. 
Thus, the compass model ($I=0$ line) is at a transition
between two phases, and four different phases meet at the isotropic 
points $J_x=\pm J_z$ (square and diamond in Fig.~\ref{fig:phd}), 
showing that both QPTs in the compass model evolve into transition 
lines between phases with different types of spin order.

Investigating the phase diagram of the CH model (Fig. \ref{fig:phd}),
perturbative approaches cannot be applied when Heisenberg interactions 
are comparable to dominant compass terms, especially when they 
frustrate each other, e.g. for $|J_x|\leq J_z$, $I<0$ and 
$|I/J_z|=O(1)$. For large negative Heisenberg couplings 
FM phases are favoured --- 
with spins being perpendicular to the $z$ axis to avoid AF compass 
couplings. Depending on the sign of $J_x$, the most favourable spin
orientation is either along $x$ ($F_x$ phase with order parameter
$M_{Fx}\equiv S^x_{(0,0)}$ for $J_x<0$) or along $y$
($F_y$ phase with $M_{Fy}\equiv S^y_{(0,0)}$ for $J_x>0$), not
found within the compass ground states manifold. By comparing
the classical energies (per site) of different phases, we
determined the classical transition lines in Fig.~\ref{fig:phd}.
For instance, the phase transitions between the $C'_z$ phase and 
both above FM phases ($F_x$ and $F_y$), were found using:
$E_0(C'_z)=-J_z$, $E_0(F_x)=J_x+2I$, and $E_0(F_y)=2I$.

\begin{figure}[t!]
\begin{center}
\includegraphics[width=8.6cm]{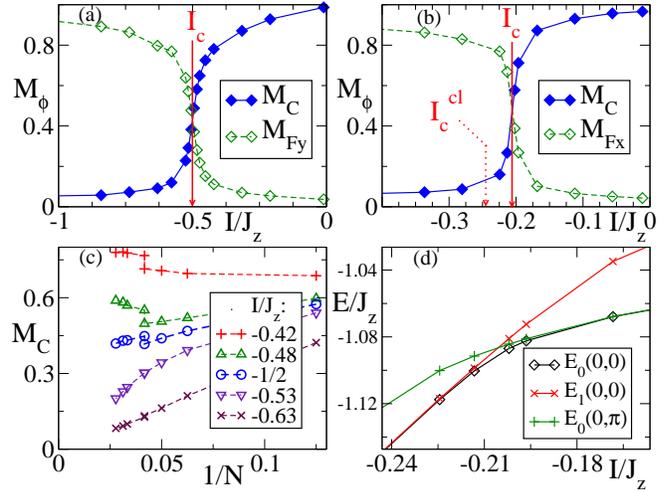}
\end{center}
\caption{Order parameters $M_\phi$ of the $C'_z$
phase ($M_C$),
and (a) of the $F_y$ phase with $\phi=\pi/10$; (b) of the $F_x$
phase with $\phi=37\pi/20$. $I_c$ indicates the transition between
$C'_z$ and either (a) $F_y$ phase, or (b) $F_x$ phase
--- in (b) its classical counterpart $I_c^{\rm cl}$ is also shown.
(c): Size-dependence of $M_C$ for values $I/J_x$ close to the
transition displayed in (a). (d): Momentum-resolved lowest energy
levels per site ($E_n(\bf{k})$ is the $(n+1)^{th}$ state of momentum
$\bf{k}$) on a $N=32$-site cluster, across the transition displayed in 
(b). The ground state energy $E_0(0,0)$ shows avoiding crossing.} 
\label{cff}
\end{figure}

Strikingly, the phase diagram of the quantum model (Fig.
\ref{fig:phd}) differs very little from that found by comparing
classical energies. First, some transition lines are not modified by 
quantum fluctuations. This occurs due to additional symmetries at 
the classical transition line. An example is the case 
of $J_x=J_z>0$ and $I>0$ discussed previously, where the
Hamiltonian is invariant under a $\pi/2$ rotation of both spins
and lattice along the $y$ axis. The first order transition point 
between two symmetry-broken compass phases, found at $J_z=J_x$ in the 
compass model \cite{Oru09} (square in Fig. \ref{fig:phd}), extends to 
a more conventional (still first order) transition line between the 
fully (classically) ordered $G_z$ and $G_x$ phases, stable for either 
$J_z > J_x$ or $J_z < J_x$.

Similarly, on the line defined by $I=-J_z/2$ and $0<J_x<J_z$
(classical transition line between $C'_z$ and $F_y$ phases), the
Hamiltonian Eq. (1) has an additional U(1) symmetry \cite{notect}, 
and therefore the corresponding QPT occurs necessarily along this line.
We have checked by considering bond correlations
$\langle\sigma_{\bf 0}^\alpha\sigma_{\bf r}^\beta\rangle$
for $\alpha\ne\beta$ that no other intermediate (e.g. spiral)
ordered phase develops in this range of parameters.
For both phases, the dependences of order parameters on $I/J_z$ and
on cluster size [shown for fixed $\phi=\pi/10$ in Fig.~\ref{cff}(a),
with size-scalings of $M_C$ in Fig.~\ref{cff}(c)] indicate a first 
order transition by the slope of order parameter which increases 
with $N$ at the transition, see Fig.~\ref{cff}(a). The scaling with 
increasing size indicates a discontinuity across the $I=-J_z/2$ point, 
whereas both order parameters are equal at this point and extrapolate 
to a finite value in the thermodynamic limit. 

We emphasize that, contrary to what happens usually in first order 
transitions, no level crossing occurs in the ground state for finite 
clusters (see also Refs. \cite{Oru09} and \cite{Vid09}), where the 
preferred spin orientation evolves continuously from the $z$ axis (in 
the $C'_z$ phase) to the $y$ axis ($F_y$), but a level crossing occurs 
between the two lowest excitations, as shown for the $C'_z-F_x$ 
transition in Fig.~\ref{cff}(d). This is related to the distinct broken 
symmetries in both phases in the thermodynamic limit:
($i$) translation symmetry breaking along $z$ in the $C'_z$ phase
results in a 2-fold degenerate ground state with ${\bf k}=(0,0)$ and 
$(0,\pi)$ momenta, while 
($ii$) the spontaneous breaking of $\mathbb{Z}_2$ symmetry in the 
$F_y$ (or $F_x$) phase results in a 2-fold ground state degeneracy, 
both states with ${\bf k}=(0,0)$ momentum 
[but with different parity $P=(-1)^{\sigma^z_{\rm tot}/2}$]. 
The crossing between the second $(0,0)$ and the lowest $(0,\pi)$
states occurs exactly at $I=-J_z/2$ in the $C'_z -F_y$ transition,
for the above symmetry reasons.

In contrast, some QPTs in the phase diagram of Fig. \ref{fig:phd},
e.g. the $C'_z-F_x$ transition (dashed line), do not
present additional symmetries at the classical level and are
affected by quantum fluctuations. The energies (per site) of
$C'_z$ and $F_x$ phase, evaluated in second order perturbation
theory, are:
\begin{eqnarray}
\label{2ord1}
E(C'_z)&=&E_0(C'_z) - \frac{J_x^2}{8J_z+4I}-\frac{I^2}{J_z-I}\,,\\
\label{2ord2}
E(F_x)&=&E_0(F_x) + \frac{J_z^2}{8J_x+12I}\,.
\end{eqnarray}
They are equal on the solid line separating both phases in
Fig.~\ref{fig:phd}. Note that numerical estimations of the
transition, with help of: ($i$) the crossing between both lowest
excitations [Fig. \ref{cff}(d)], and ($ii$) the increase/decrease of
$M_{Fx}$ and $M_C$ order parameters [Fig.~\ref{cff}(b)] are not
only consistent between themselves, but also in very good
agreement with Eqs. (\ref{2ord1}) and (\ref{2ord2}).

Quantum fluctuations also shift somewhat the transition line between 
$G_z$ and $C_x$ phases (for $J_x<-J_z$). Contrary to intuition, 
their contribution is here larger when neighbouring spins are aligned.
Thus the FM phase is stabilized by them over the $C'_z$ phase
(at the $F_x$-$C'_z$ transition), and the $C_x$ phase over
the $G_z$ phase (at the $C_x$-$G_z$ transition).

\begin{figure}[t!]
\begin{center}
\includegraphics[width=8.2cm]{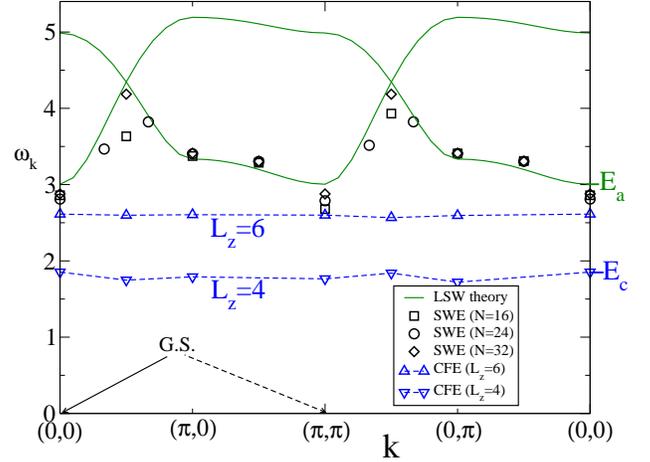}
\end{center}
\caption{The two types of excitations in the
Compass-Heisenberg model: spin-wave excitations (SWE) with the lowest 
energies for clusters of $N=16, 24, 32$ sites (squares, circles and 
diamonds); the LSW dispersion Eq. (\ref{omega}) shown by solid lines, 
with minima $E_a$ at ${\bf k}=(0,0)$ and 
${\bf k}=(\pi,\pi)$), see Eq. (\ref{ea}); and column-flip 
excitations (CFE) with energies $E_c$ (triangles) for two values of 
$L_z=4, 6$ (with $L_x=4$), see Eq. (\ref{ec}). 
Parameters $I=0.1J_c$ and $\phi=3\pi/20$ 
are corresponding to $G_z$ phase in Fig. 1. 
}
\label{swc}
\end{figure}

Next we analyze low-energy excitations for finite clusters
which depend on the interaction parameters in a remarkable way.
There are two fundamentally different types of excitations
for the CH model Eq. (\ref{ham}): 
($i$) {\it spin waves\/}, 
i.e., coherent propagation of single spin flips, and
($ii$) {\it column flips\/}, 
where all spins of a column are flipped.
First, we employ linear spin-wave (LSW) theory to estimate the
dispersions of spin waves in various ordered phases with an adapted 
vacuum state (for $C$- or $G$-like phases, a canonical transformation
\cite{notect} allows one to use a single type of bosons as for FM
phases). With a Bogoliubov transformation, one finds 
spin-wave dispersion
for the $G_z$ phase (for $|J_x|<J_z$ and $I>0$) 
\begin{equation}
\label{omega}
\omega_{\bf{k}}=2\sqrt{(2J_z+J_{\bf{k}}+4I)^2
                -(J_{\bf{k}}+I_{\bf{k}})^2} \,,
\end{equation}
with $J_{\bf{k}}=J_x\cos k_x$ and $I_{\bf{k}}=2I(\cos k_x+\cos k_z)$. 
The ground state is 2-fold degenerate in the thermodynamic limit with 
momenta ${\bf k}=(0,0)$ and $(\pi,\pi)$ 
as indicated in Fig. \ref{swc}, and there are two spin-wave
branches with dispersion $\omega_{\bf{k}}$ and $\omega_{\bf{k}+\bf{Q}}$,
respectively. 
The agreement between these two branches and the lowest spin-wave 
excitation energies obtained for finite clusters with $N\le 32$ sites
(identified as excitations to the lowest states with $P=-1$) is 
satisfactory, see Fig. \ref{swc}. 
The minimum of spin-wave dispersion $E_a$ or anisotropy energy, e.g.
\begin{equation}
\label{ea}
E_a= 2\sqrt{(2J_z-J_x+4I)^2-(J_x+4I)^2}
\end{equation}
in the $G_z$ phase with excitations displayed in Fig. \ref{swc}, 
is generally finite in an ordered phase, and vanishes along 
a transition line to another phase (except 
for the $I=0$ line). 
An example is the $G_z - G_x$ transition line, where numerics and the 
LSW theory indicate a gapless spectrum, in contrast with the isotropic
three-dimensional CH model which has a finite gap \cite{Kha01}. The 
transition lines characterized by spin-wave softening differ from the 
$I=0$ line --- there, e.g. at the $G_z - C'_z$ transition, spin waves 
remain gapped and a softening of $(2^L-2)$ columnar excitations occurs.

For finite clusters and for small enough $I/J_z$
excitations to other compass states 
have lower energies than the spin waves. 
The simplest ones of these excitations, called \textit{column flip}, 
consist of reversing all spins of a \textit{single} column ---
these excitations are fundamentally different from spin waves which
occur for Heisenberg systems with long range order. The resulting 
excitation energy is mainly due to Heisenberg couplings between each 
spin of the column and its two horizontal neighbours, and is given by 
\begin{equation}
\label{ec}
E_c \simeq 4L_z|I|, 
\end{equation}
thus increasing linearly with column size. 
The smallness of the coupling $J_{\rm col}$ Eq. (\ref{Jcol}) between 
columns gives rather weak $\bf{k}$-dependence of these excitations, 
as shown in Fig. \ref{swc} for $L_z=4$ and fixed $L_x=4$. This coupling 
decreases with increasing column length, being
$J_{\rm col}=2.3\times 10^{-3}$ for $L_z=4$ and 
$J_{\rm col}=1.2\times 10^{-4}$ for $L_z=6$ (both with $\phi=3\pi/20$) 
--- thus almost no dispersion is seen for $L_z=6$. These excitations are
shifted above the spin-wave ones with increasing size $L_z$ and play 
no role in the thermodynamic limit. 

\begin{figure}[t!]
\begin{center}
\includegraphics[width=8.2cm]{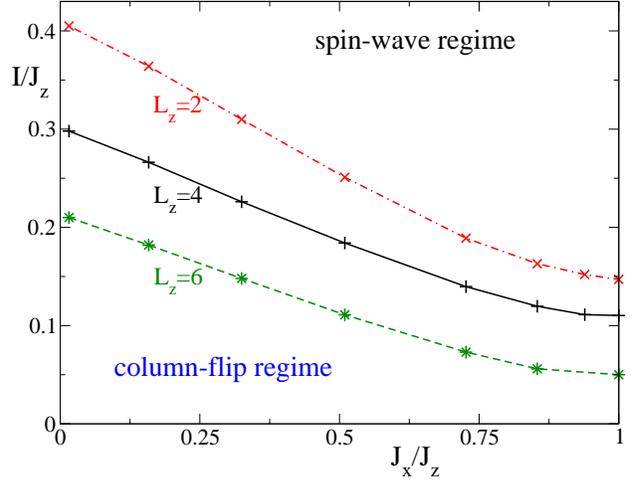}
\end{center}
\caption{
Two distinct regimes of low energy excitations
for $J_x<J_z$ \cite{notejz}: 
column-flip regime with $E_a > E_c$ (lower part) and 
spin-wave regime with $E_a < E_c$ (upper part), as obtained for 
$\phi=3\pi/20$ and for $L_x\times L_z$ clusters with: 
$L_x=4$ and $L_z=2,4,6$. 
For each system size, $E_a$ and $E_c$ are calculated separately. 
}
\label{cfe}
\end{figure}

A comparison of the column-flip excitations with spin waves is of
importance for finite clusters since such column flips could be used 
for fault-tolerant quantum computing, where a qubit would be encoded 
in the orientation of a given spin column. A criterion for the use of 
such a device is that spin-wave excitation energies remain above those 
of column flips, i.e., $E_a > E_c$, which defines a 
\textit{column-flip regime}. In Fig. \ref{cfe} we present regimes of 
these distinct excitations obtained for three clusters in the case of 
AF interactions, with fixed number of $L_x=4$ columns and increasing 
column length $L_z=2,4,6$. On the one hand, the size of columns and the 
anisotropy ratio $J_x/J_z$ determine the range of the column-flip 
regime, which requires large coupling anisotropy (small $J_x/J_z$) and 
sufficiently short columns (since $E_c\propto L_z$). In the perturbative 
regime of small $I/J_z$ and $J_x/J_z$ one finds the transition between 
these two distinct regimes at
\begin{equation}
\label{cross}
L_z I=\sqrt{J_z(J_z-J_x)} +f_L\,.
\end{equation}
The factor $f_L$, accounting for finite-size corrections to the 
spin-wave dispersion, vanishes in the thermodynamic limit; but
remarkably, even in the vicinity of the isotropic point 
($J_x\leq J_z$), these corrections allow compass excitations to be 
the lowest ones for small enough $L_z|I|$. On the other hand, $L_z$ 
must be large enough for column-flip excitations of different columns 
to be sufficiently far from one another in Hilbert space, so that 
qubits remain well protected against local fluctuations and noise 
\cite{Dou05}. 
Thus, for given values of $\{J_x,J_z\}$ couplings the system size must 
correspond to a compromise between all constraints above, in order to 
define correctly qubits with help of these column-flip excitations.

In contrast to frequently proposed quantum computing schemes
\cite{Dou05,Compu}, here a column-flip switches between quasi-degenerate
eigenstates (split by an energy $\propto J_{\rm col}$); thanks to their
columnar character the fault tolerance of the compass model persists, 
unaltered by perturbating Heisenberg interactions --- imperfections in 
switches should not harm information storage more than usual decoherence
sources. 

One may also wonder whether the features discussed above are a 
consequence of the particular nature of perturbing interactions. 
We can for instance consider, instead of Heisenberg couplings, 
XY-type couplings --- or rather XZ-type within our notation, 
i.e., introducing $I_{xz}(\sigma_i^x \sigma_j^x+\sigma_i^z \sigma_j^z)$ 
for each pair of nearest neighbours, such that the $\sigma_i^y$ spin 
components do not appear in the Hamiltonian anymore. 
One finds that the order induced by perturbations at even infinitesimal 
$I_{xz}$ persists since e.g. the neighbouring columns in column-ordered 
compass states are again coupled at first order in perturbation. 
The global phase diagram would mainly differ from that found for the CH 
model (presented in Fig. \ref{fig:phd}) by the absence of $F_y$ phase, 
and the transition lines would not be affected by quantum fluctuations 
at all. Nevertheless, the column flips will also be the lowest energy 
excitations for sufficiently small perturbation amplitude $I_{xz}$ and 
system size, allowing here again for a possible design of a quantum 
computation scheme.

Summarizing, we have shown that the macroscopic $2^L$ ground state
degeneracy of the anisotropic $L\times L$ compass model is lifted by
infinitesimal Heisenberg interaction $I$. The Compass-Heisenberg model 
has a rich phase diagram --- for small $|I|$ long-range order develops 
from a pair of compass states, and the remaining compass states are 
split off by an energy $E_c\sim 4 L|I|$. Thus the spin waves are the 
lowest energy excitations in a large system. 
For nanoscale structures of length $L$, however, the sequence of 
excited states can be reversed, with quasi-degenerate compass states 
being pushed below the spin-wave excitations, provided the product 
$L|I|$ is small enough. In this way decoherence 
of column-flip excitations by decay via 
spin waves could be avoided in quantum computation applications.

\acknowledgments
We thank B. Dou\c{c}ot and G. Khaliullin for insightful discussions. 
F.T. acknowledges partial support by the European Science Foundation 
(Highly Frustrated Magnetism network, Exchange Grant 2525).
A.M.O. acknowledges support by the Foundation for Polish Science 
(FNP) and by the Polish Ministry of Science and Higher Education 
under project N202 069639.


\end{document}